# Footstep Recognition As People Identification : A Systematic Literature Review

Arif Rachmat

*Universitas Islam Negeri Datokarama Palu*

Palu, Indonesia

a@uindatokarama.ac.id

*Abstract* — Footstep recognition is a relatively new biometric which aims to discriminate people using walking characteristics. There are several feature and technology have been adopted in various research. This study will attempt to show a comparative technology and feature which is offered each previous related works. We performed a broad manually search to find SLRs published in the time period 1st January 2006 to 30th November 2018. Our broad search found 12 SLRs articles refer to 3 similar technology and 5 cluster feature. In over time, the number of published footstep recognition has increased, especially in conference publications. The differences in footsteps can be known from the power spectral density of sounds and vibrations generated by footsteps. Every footstep of the human has a certain density of frequency, either from density of sounds or vibrations generated. To improve accurately of the result, this paper suggests furthermore research to combining several measuring sensor and data processing method.

## I. INTRODUCTION

Identity verification is a crucial problem in a security system. Common biometric properties such as iris [1], fingerprint [2], palm-print [3] and voice [4] have been used to provide additional security. The biometric properties can be divided into two categories: the physiological characteristics (e.g. iris, fingerprint and palm-print) and the behavioral characteristics (e.g. voice and keystroke). Physiological characteristics are relatively stable and unique across a large section of the population; behavioral traits, on the other hand, have some physiological basis, but also reflect a person's psychological makeup and environmental variations. Footsteps are a universally available signal that humans can frequently classify correctly.

The important benefit of footsteps over other biometric properties is that footstep signals can be collected imperceptibly with no personal cooperation, and do not reveal any identity to other humans such as a face or a voice does. Accurate footstep analysis would be useful in various applications such as home security service, surveillance and understanding of human action. Using footsteps, one can easily identify an acquaintance, but constructing an identity verification system based on them remains a challenging problem: footsteps, as a main kind of behavioral traits, not only reflect a person's physiological basis, but also depend on his or her psychological makeup, footwear and floor on which he or she is walking. Furthermore, since footsteps are usually accompanied with noisy and environmental sounds in practical applications, it is difficult to extract stable footstep features from the mixed sounds. Three main issues limit footstep identification performance and applications [5]: Sensitive to changes of footwear and floor: As behavioral traits, footsteps greatly depend on the types of footwear and floor. However, in practice, the variations of footwear are inevitable. Sensitive to the person's psychological makeup: It is well known that footsteps reflect a person's psychological makeup, as such , the similarity examinination in different psychological states yields different footsteps. It is unrealistic for examiners to keep their psychological states unchanged at all times due to sensitivity to noisy and environmental sounds: In practice, footsteps are usually accompanied with noisy and environmental sounds, unfortunately, there is no perfect acoustic feature extraction method which can extract stable features from the footsteps mixed with noisy or environmental sounds.

To extend understanding of footstep recognition systems, the focused of this paper is to conduct a literature review using systematic literature review (SLR) method. SLR is a secondary study to collect data from relevant primary studies [3]. SLR could assist to find a solution by performing a review on the previous relevant research. The objective of the study is to understand the proposed feature and sensor technology of a footstep recognition system. This understanding can be used as a base for developing a platform of footstep recognition system. To obtain a comprehensive result, this study undertook the research on several published literature from popular database journal e.g. IEEE Xplore, Scopus, SpringerLink, ScienceDirect, and ACM from 2006 to 2018.

This paper consists of four sections. Section 2 gives an explanation the systematic literature review as the methodology of this study. Section 3 describes the result of the review and the answer to the research question. Section 4 presents the conclusion of the research and suggestions for future works.

## II. RESEARCH METHODOLOGY

### A. Systematic Literature Review

Systematic Literature Review (SLR) is a secondarydata collection method to map, identify, critically evaluate, consolidate, and collect the results of relevant primary studies on a certain research topic [6]. SLR becomes a standard method to obtain an answer by performing a literature review based on the previous relevant studies. The purpose of performing SLR is to summarize the previous research, to identify the gap which needed to be fulfilled between the previous and the current research, to produce a coherent report/synthesis, and to make a research framework.

The purpose of a literature study on this research is to further understand the feature, technology, method, and platform of a survey data collection system. To obtain a comprehensive result, the study undertook the research on several published literature from popular database journal e.g. IEEE Xplore, ScienceDirect, and ACM from 2006 to 2018.

SLR is a process to identify, evaluate, and interpret the research results to answer the determined research question. SLR consists of several stages. They are determining the research question, selecting the corresponding research, mining the required data, analyzing and describing the discovery.

The SLR is a structured and systematic approach for the identification, selection and synthesis of recent literature relevant to the research question in hand. SLRs are a means of aggregating knowledge about a software engineering topic or research question

### B. Research Question

The purpose of the research question is to maintain the focus of the literature review. This condition leads to more convenient mining process for the required data. Table 1 shows the research questions for this research.

TABLE I. RESEARCH QUESTION

| ID | Research Question | Motivation |
|---|---|---|
| RQ1 | Which features are proposed by footstep recognition system? | Identify the type of sensor has been used for footstep recognition system to capture data |
| RQ2 | Which technologies are support footstep recognition system? | Identify the proposed feature by footstep recognition systems |
| RQ3 | How accurate the method has been used? | Identification successful of using the device and the process data methodology |

### C. Research Finding

To answer the above research questions, the study conducted a search on the published research paper in the popular database journal with specific string for searching, from November 25, 2018 to December 8, 2018. The string is "footstep" AND (identify OR identification OR recognition). Table 2 shows the results of the search.

TABLE II. THE RESULT FINDING OF RELATED STUDY

| ID | Database Journal | Number of Articles |
|---|---|---|
| 1 | IEEE Xplore | 51 |
| 2 | ACM Digital Library | 61 |
| 3 | ScienceDirect | 621 |
| 4 | SpringerLink | 125 |
| 5 | Scopus | 94 |
| Total | | 952 |

The study applied several inclusion and exclusion criteria to filter the article for further exploration. Table 3 shows the criteria.

TABLE III. INCLUSION AND EXCLUSION CRITERIA

| | Criteria |
|---|---|
| Inclusion Criteria | • I1 – Articles relate to feature, and technique/method for footstep recognition<br>• I2 – Articles written in English<br>• I3 – Articles can be accessed in full-text |
| Exclusion Criteria | • E1 – Similar articles from different database journal<br>• E2 – Articles in a book, proceedings, conference review, or white paper<br>• E3 – Articles not related to people recognition or people identification or biometric.<br>• E4 – Footstep recognition without human detection. |

Figure 1 displays the filter stages performed in accordance with the specified criteria, from the filter for the title/metadata, abstract, authorized access, and full-text review.

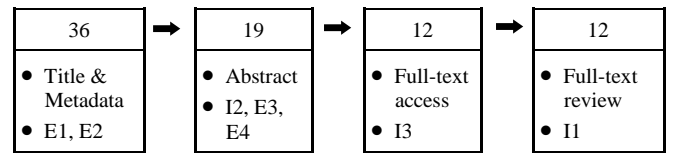

Fig. 1. The Filtering Process

In the first filtering, the criteria E1 & E2 applied to the title and metadata section reduced the number of papers from 952 to 36. The second filtering, criteria I2, E3 and E4 applied to the abstract section reduced the number of papers to 19. The Third filtering, criteria I3 (authorized access) applied makes the number of papers to 12. The last filtering, the criteria I1 applied to full-text review makes the number of papers to 12.

After applying the above inclusion and exclusion criteria, several articles, which published in periodic journal or conference proceeding, satisfy the criteria and can be utilized as the main reference for SLR. Table 4 shows the result of the filtering process.

TABLE IV. THE RESULT OF FILTERING PROCESS

| Index | Publication Media | Articles Found |
|---|---|---|
| 1 | Periodic Journal Q1 | 7 |
| 2 | Periodic Journal Q2 | 3 |
| 3 | Periodic Journal Q3 | 0 |
| 4 | Periodic Journal Q4 | 2 |
| Total | | 12 |

## III. RESEARCH RESULT

### A. Sensor technology

For collecting footstep data, there are various technologies can be used. Based on literature review, it can be grouped into several methods based on data-type : pressure signal, seismic signal, and acoustic signal. Table 5 describes the sensor technology and the appropriate literature.

TABLE V. SENSOR TECHNOLOGY

| Index | Sensor | Literature |
|---|---|---|
| 1 | Piezoelectric & EMFi (pressure signal) | [10] [9] [7] [8] [12] [11] [13] |
| 2 | Geophone (seismic signal) | [15] [16] [14] |
| 3 | Acoustic sensor (acoustic signal) | [5] [17] |

Piezoelectric and electro mechanical film (EMFi) have similar methods in capturing footstep shifting, which the sensor usually placed on floors or mat, it senses the pressure changes affecting its surface and provides footstep profiles of the walking person as an input to the identification system [10], furthermore in order to detect the shape and position of the foot [9].

In order to detect the person or vehicles, mostly uses three-component seismic velocity transducer [14]. Movement over the ground and this disturbance which generate from the soil disturbance propagates away from the moving source is captured by geophone seismic sensor as seismic waves raw data [15] [16] [14]. Geophones are most effective when buried, thereby making them covert.

Footstep-identification methods can use acoustic and psycho-acoustic parameters or footstep pressure dynamics captured from people walking over an instrumented sensing area as footstep features. They're sensitive to change in footwear, floor surface, and the walker's psychological state, so in these methods, the footwear, floor, and walking manner (single person in general psychological makeup) are limited [5].

*B. Footstep proposed features*

There are several feature are proposed, and grouped by some categories. Table 6 shows the list of feature proposed for footstep identification.

TABLE VI. PROPOSED FEATURE

| Index | Proposed Feature | Literature |
|---|---|---|
| 1 | Object detection | [15] [14] [12] |
| 2 | Person identification | [10][5][8][17][16][13] |
| 3 | Person characteristic | [9] [7] [11] |
| 4 | Person behavior | [9] [7] [8] [11][12] |
| 5 | Intrusion detection | [14] [15] |

In this case, object detection is ability to determining by comparing the time difference between impulsive events created by that target. In humans, animals or vehicles, the variance in time the difference between impulses is usually small over short periods of time so it is important to analyze cadence in a certain time window [15], it has been proposed for a number of different applications, including security, surveillance, tracking persons in an area and recognizing human behavior [12].

Person identification in intelligent environments has become a major research issue, which is related to the biometric identification domain [10]. The behavioral characteristics of a walking person are used to model the person's identity. The biometric person identification system is based on sensor measurements achieved from a pressure-sensitive floor on this study.

An alternative way to representing people characteristic is how the systems can regarding the type of footwear employed, persons are free to walk with different types of footwear such as shoes, trainers, boots, flip-flops, barefoot and even high heels [9]. Also, people are knowing if carry weights such as office bags.

Persons behavior can be known by learning and tracking human directions movement frequency which they often, spend different times in certain places [15].Thus is certainly very useful for retail companies who want to know their consumers interest places. Footsteps also have some potential applications in the smart home environment where footstep sensors are installed to determine the position of a person in a room or to recognize human behavior and interact with users [12].

The principle behind the intrusion detection is generation of the succession of impacts when the person, animal or vehicle move over the ground and this disturbance which generate from the soil disturbance propagates away from the source as seismic waves [14]. There is a growing need for perimeter detection technologies that can be deployed in remote locations over long distances. These needs are growing not only in military environments, but also in homeland defense applications where persistent border security is of increasingly high priority [15].

*C. Classifier and accuracy*

The accuracy of each classifier method used is very dependent on the level of quality of the measuring instrument and research on the measurement algorithm. Table 7 show the list of claimed accuracy.

TABLE VII. ACCURACY RESULT

| Index | Claimed Accuracy (%) | Literature |
|---|---|---|
| 1 | 96 - 100 | [13], [15] |
| 2 | 91 - 95 | [5], [7], [8], [10] |
| 3 | 86 - 90 | [9], [12], [16] |
| 4 | 81 - 85 | [17] |
| 5 | N/A | [11][14] |

IV. CONCLUSION

The objective of the Systematic Literature Review is to find and understand the sensor technology and proposed feature by conducting a search on the previously published research between 2006 and 2018 on several popular database journals. The study find research related to footstep recognition. Object detection, person identification, person behavior, intrusion detection are offering feature found in the result. For sensor technology can divide by 3 group of methods : pressure sensor, seismic sensor and acoustic sensor. Pressure sensor is the most widely used as sensor technology to collection footstep data in previous research. Combining sensor was not found in this searching. Integration several type of sensors and methods may can applied to develop a more accurate result for footstep systems recognition.